**Αργή και γρήγορη σκέψη και εκπαίδευση στις φυσικές επιστήμες**

*Δρ. Δημήτριος Γουσόπουλος, Εκπαιδευτικός*

**Περίληψη**

Τα τελευταία χρόνια υπάρχουν ολοένα και περισσότερες ενδείξεις ότι ακόμη και μετά από μία διδασκαλία που έχει σχεδιαστεί για να αντιμετωπίσει τις μαθησιακές δυσκολίες που υπαγορεύει η βιβλιογραφία, πολλοί μαθητές αποτυγχάνουν να δημιουργήσουν τις σωστές συλλογιστικές οδούς που συνδέουν τις θεμελιώδεις αρχές και οδηγούν σε αιτιολογημένες προβλέψεις. Παρόλο που οι μαθητές διαθέτουν τις απαιτούμενες γνώσεις και δεξιότητες, συχνά βασίζονται σε μια ποικιλία διαισθητικής συλλογιστικής που τους οδηγεί σε λάθος συμπεράσματα. Στην παρούσα εργασία επιχειρούμε, με βιβλιογραφική επισκόπηση, να διερευνήσουμε τη συλλογιστική των μαθητών σε προβλήματα Φυσικών Επιστημών μέσω ευρετικών - αναλυτικών διαδικασιών σκέψης (Σύστημα 1 - Σύστημα 2). Το σύστημα 1 λειτουργεί αυτόματα και γρήγορα με λίγη ή καθόλου προσπάθεια, ενώ το Σύστημα 2 εστιάζει την προσοχή στις απαιτητικές νοητικές δραστηριότητες που το απαιτούν και είναι αργό βασισμένο σε κανόνες.

**Λέξεις-Κλειδιά:** εκπαίδευση στις φυσικές επιστήμες, ευρετικές, συστημα 1, σύστημα 2, διερεύνηση δυσκολιών στις φυσικές επιστήμες

**Thinking fast and slow and science education**

*Dr. Dimitrios Gousopoulos, science teacher*

**Abstact**

In recent years there has been growing evidence that even after teaching designed to address the learning difficulties dictated by literature, many physics learners fail to create the proper reasoning chains that connect the fundamental principles and lead to reasoned predictions. Even though students have the required knowledge and skills, they are often based on a variety of intuitive reasoning that leads them to wrong conclusions. This paper studies students' reasoning on science problems through heuristic - analytical thought processes (System 1 - System 2). System 1 operates automatically and quickly with little or no effort and no sense of voluntary control, while System 2 focuses on the demanding mental activities that require it and is slow based on rules.

**Key-Words:** science education, heuristics, system 1, system 2, investigating students' difficulties in science

**Εισαγωγή**

Πολλοί μαθητές, στα μαθήματα φυσικών επιστημών, συχνά αποτυγχάνουν να δημιουργήσουν τις κατάλληλες συλλογιστικές οδούς που συνδέουν τις θεμελιώδεις αρχές και παράγουν επιφανειακές απαντήσεις στις ερωτήσεις και τα προβλήματα που αντιμετωπίζουν (Meltzer & Thornton, 2012). Η ερευνητική – εκπαιδευτική κοινότητα έχει αναπτύξει στοχευμένες διδασκαλίες για την αντιμετώπιση των εννοιολογικών και συλλογιστικών δυσκολιών των μαθητών. Ωστόσο, σε πολλά πλαίσια, αυτές οι παρεμβάσεις δεν οδήγησαν σε σημαντικές βελτιώσεις στην επίδοση των μαθητών, παρόλο που οι τελευταίοι κατέχουν τις απαιτούμενες γνώσεις και δεξιότητες (Kryjevskaia et al., 2012). Πιο συγκεκριμένα, ένα αναδυόμενο σύνολο στοιχείων υποδηλώνει ότι οι μαθητές εγκαταλείπουν τον αναλυτικό τρόπο προσέγγισης των προβλημάτων και βασίζονται σε έναν πιο διαισθητικό τρόπο σκέψης που συχνά τους οδηγεί σε λάθος απαντήσεις. (Kryjevskaia et al., 2014)

Τα τελευταία χρόνια, οι ερευνητές προσπάθησαν να ρίξουν φως στον διαισθητικό τρόπο που σκέφτονται οι μαθητές χρησιμοποιώντας τις θεωρίες περί των δύο συστημάτων του νου. Για παράδειγμα, έρευνες έχουν διερευνήσει και αναλύσει τις ευρετικές που χρησιμοποιούνται από τους μαθητές για την ταξινόμηση χημικών ενώσεων (McClary & Talanquer, 2011 ; Talanquer, 2014), ενώ άλλες έχουν εξετάσει τη διαισθητική σκέψη των μαθητών σε θέματα φυσικής όπως η δύναμη (Wood et al., 2016) και άνωση (Gette et al., 2018). Οι θεωρίες περί των δύο συστημάτων του νου έχουν δώσει μια εικόνα για το πώς οι άνθρωποι συλλογίζονται υπό συνθήκες περιορισμένου χρόνου, γνώσης και υπολογιστικής ισχύος (Shah & Oppenheimer, 2008; Kahnemann, 2011), οι οποίες είναι παρόμοιες με αυτές που αντιμετωπίζουν οι μαθητές μας σε μια τάξη φυσικής.

Οι εκπαιδευτικές έρευνες εντοπισμού των βασικών μηχανισμών που οδηγούν στα παρατηρούμενα μοτίβα σκέψης είναι ελάχιστες, παρά το γεγονός ότι οι γνωστικοί επιστήμονες έχουν αναπτύξει διάφορα μοντέλα ερμηνείας της ανθρώπινης σκέψης και λήψης αποφάσεων.

## Περί των δύο συστημάτων του νου

Είναι μια κοινή προσδοκία ότι, μετά από κατάλληλη διδακτική παρέμβαση, οι μαθητές θα κατασκευάσουν συνειδητά και συστηματικά αλυσίδες συλλογισμών που ξεκινούν από την καθιερωμένες επιστημονικές αρχές και οδηγούν σε αιτιολογημένες προβλέψεις. Όταν η απόδοση των μαθητών στις εξετάσεις μαθήματος δεν αποκαλύπτει τέτοια μοντέλα, συχνά θεωρείται ότι οι μαθητές είτε δεν κατέχουν την κατάλληλη κατανόηση της σχετικής επιστημονικής γνώσης είτε δεν είναι σε θέση να κατασκευάσουν τέτοιες αλυσίδες αιτιολογικών συλλογισμών (Heckler,2011). Η έρευνα στους τομείς της σκέψης και των συλλογισμών φαίνεται, από την άλλη πλευρά, να προτείνει ότι στις περισσότερες περιπτώσεις η σκέψη ακολουθεί μονοπάτια τελείως διαφορετικά σε σχέση με αυτά που αναφέρθηκαν προηγουμένως. Η θεωρία περί των δύο συστημάτων του νου προτείνει ότι υπάρχουν δύο διαφορετικές διαδικασίες που εμπλέκονται σε πολλές γνωστικές εργασίες, γνωστές ως Σύστημα 1 (περιέχει τις ευρετικές διαδικασίες) και Σύστημα 2 (περιέχει τις αναλυτικές διαδικασίες). Το πρώτο σύστημα υποστηρίζει συλλογισμούς που είναι γρήγοροι ,διαισθητικοί και αυτόματοι, ενώ το δεύτερο σύστημα είναι αργό και βασισμένο σε κανόνες. Σε πολλές περιπτώσεις αυτά τα δύο Συστήματα του νου παράγουν διαφορετικές απαντήσεις. Ο Evans (2006) μέσω ενός διαγράμματος, που φαίνεται στην παρακάτω εικόνα 1, παρέχει ένα οπτικό βοήθημα για την παρακολούθηση και την κατανόηση των διαφόρων συλλογιστικών διαδρομών που προκύπτουν από την αλληλεπίδραση των δύο Συστημάτων.

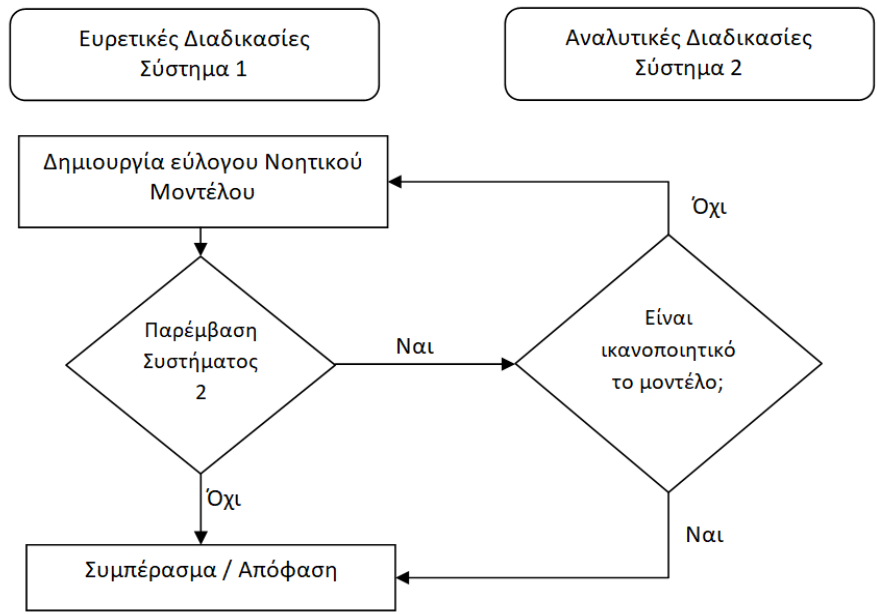

Εικόνα 1

Όταν ένα άτομο έρχεται αντιμέτωπο με μια συγκεκριμένη κατάσταση ή ένα πρόβλημα προς επίλυση, το Σύστημα 1 άμεσα και υποσυνείδητα αναπτύσσει ένα νοητικό μοντέλο της κατάστασης με βάση τις προηγούμενες γνώσεις και εμπειρίες, συναφή στοιχεία και άλλους παράγοντες. Αυτό το «πρώτο διαθέσιμο» νοητικό μοντέλο συχνά αντιπροσωπεύει μια γρήγορη και υποσυνείδητη προσπάθεια εκ μέρους του ατόμου να παράγει ένα συνεκτικό και αξιόπιστο τρόπο να προσεγγίσει τη δοθείσα κατάσταση ή πρόβλημα. Σύμφωνα με τον Evans, ο άνθρωπος δημιουργεί ένα μοντέλο κάθε φορά (singularity principle). Το μοντέλο που προκύπτει από τις ευρετικές διαδικασίες του νου βασίζονται στις πιο πρόσφατες γνώσεις και πεποιθήσεις του ατόμου που προκαλείται από τα συμβατά με τους τρέχοντες στόχους του ατόμου (relevance principle). Μόλις το πρώτο διαθέσιμο νοητικό μοντέλο αναπτύσσεται, γίνεται διαθέσιμο για έλεγχο από το πιο αυστηρό και αναλυτικό Σύστημα 2. Ωστόσο, εάν το άτομο αισθάνεται εμπιστοσύνη στο πρώτο διαθέσιμο νοητικό μοντέλο, ενδέχεται η αναλυτική διαδικασία να παρακαμφθεί εξ ολοκλήρου. Αν συμβεί αυτό, τότε το πρώτο διαθέσιμο μοντέλο παράγει την τελική απάντηση. Αυτή η άμεση μετάβαση από την πρώτη αίσθηση στην τελική απάντηση (χωρίς απαραίτητα την παρέμβαση του Συστήματος 2) επικρατεί στις καθημερινές μας δραστηριότητες. Σε γενικές γραμμές το Σύστημα 1 είναι αρκετά αποδοτικό και ακριβές στην παροχή γρήγορων αξιολογήσεων γνωστών καταστάσεων. Επίσης ακόμα κι αν το Σύστημα 1 προτείνει ένα λάθος συμπέρασμα ή μια λανθασμένη δράση σε κάποιον τομέα της καθημερινότητας, συχνά δεν είμαστε σε θέση να συνειδητοποιήσουμε ρητά την αποτυχία αυτών των συμπερασμάτων, αφού δεν δεχόμαστε άμεση ανατροφοδότηση για τα λάθη μας. Λόγω αυτού δεν αποτελεί έκπληξη ότι μαθαίνουμε να εμπιστευόμαστε τη διαίσθησή μας, η οποία μας φαίνεται να είναι αποδοτική μέσα καθημερινή μας ζωή. Τα προηγούμενα έχουν σαν αποτέλεσμα οι αρχάριοι μαθητές φυσικής να μεταφέρουν την εμπιστοσύνη τους στις διαισθητικά ελκυστικές σκέψεις από το πλαίσιο της καθημερινότητας στο πλαίσιο της επιστήμης και τελικά να μην χρειάζονται συχνά τη ρητή και αυστηρή επικύρωση της σκέψης τους. Για να εντοπιστεί ένα λάθος της διαισθητικής σκέψης θα πρέπει να εμπλακεί το Σύστημα 2 και να τεθεί σε επαγρύπνηση. Αν η αναλυτική σκέψη δεν μείνει ικανοποιημένη με την τρέχουσα απάντηση του Συστήματος 1, τότε η διαδρομή λογικής επιστρέφει στο Σύστημα 1 για να παράγει ένα νέο νοητικό μοντέλο προς εξέταση (εκ μέρους του Συστήματος 2). Είναι σημαντικό να τονίσουμε ότι ακόμα και αν εμπλακεί το Σύστημα 2 έχει τις δικές του μεροληψίες. Ως εκ τούτου, η εμπλοκή των αναλυτικών διαδικασιών δεν διασφαλίζει αναγκαστικά την ανίχνευση ενός ελαττώματος στην πορεία σκέψης ή την παραγωγή απάντησης βασισμένης σε λογικά επιχειρήματα. Για παράδειγμα, οι άνθρωποι τείνουν να δημιουργούν συνεκτικά νοητικά μοντέλα αναζητώντας επιβεβαιωτικά στοιχεία για να υποστηρίξουν αυτό το πρώτο διαθέσιμο νοητικό μοντέλο. Σε αντίθεση με την ορθή επιστημονική σκέψη, τα άτομα συχνά αποτυγχάνουν να αναζητήσουν αυθόρμητα για εναλλακτικά νοητικά μοντέλα ή για στοιχεία τα οποία θα μπορούσαν, ενδεχομένως, να τροποποιούν τις αρχικές προβλέψεις τους. Επομένως, ακόμη και όταν η αναλυτική διαδικασία έχει εμπλακεί, η πρώτη εντύπωση, η διαισθητική απάντηση συχνά εμφανίζεται ως η τελική απάντηση (Frankish,2010; Johnson-Laird,2006; Tversky & Kahneman,1973; Thompson, Turner & Pennycook,2011; Thompson, Evans & Campbell,2018).

Πολλές από τις διαισθητικές διαδικασίες του Συστήματος 1 μπορούν να θεωρηθούν σαν στρατηγικές συντόμευσης οι οποίες ονομάζονται ευρετικές και οι οποίες μειώνουν τις πληροφορίες και τα στοιχεία που επεξεργάζεται ο νους. Σε γενικές γραμμές, οι ευρετικές απλουστεύουν τη συλλογιστική ενός ατόμου μειώνοντας τον αριθμό των στοιχείων που χρησιμοποιούνται για τη λήψη απόφασης ή παρέχοντας σιωπηρούς κανόνες για το πώς και πού να αναζητήσουν πληροφορίες, πότε να σταματήσουν την αναζήτηση και τι να κάνουν με τα αποτελέσματα (Gilovich, Griffin & Kahneman,2002; Kahneman,2011; Shah & Oppenheimer,2002). Στη συνέχεια, θα αναφερθούμε σε μερικές από τις πιο γνωστές ευρετικές διαδικασίες.

### Ευρετικές διαδικασίες και εκπαίδευση στις φυσικές επιστήμες

#### Γνωστική ευκολία

Οι πληροφορίες που παρουσιάζονται σε ένα πρόβλημα ή μια κατάσταση δεν επεξεργάζονται όλες με την ίδια ευκολία από το νου. Για τους αρχάριους, τα σαφή χαρακτηριστικά γνωρίσματα είναι πιο εύκολα προσπελάσιμα από ότι τα πιο σιωπηρά. Η γνωστική ευκολία συνδέεται με το χρόνο επεξεργασίας των εμπλεκόμενων με το πρόβλημα στοιχείων και σύμφωνα με αυτή: όσο πιο γρήγορα ο νους επεξεργάζεται μια ιδέα, τόσο περισσότερο "βάρος" δίνεται σε αυτή κατά τη δημιουργία μιας συλλογιστικής πορείας (Morewedge & Kahneman,2010; Oppenheimer,2008).

Τα στοιχεία που είναι ευκολότερο να παρατηρηθούν και να επεξεργαστούν από ένα μαθητή αναμένεται να επηρεάσουν την απάντησή του σε ένα δοσμένο πρόβλημα, ιδιαίτερα αν τέτοια στοιχεία μπορούν να συσχετιστούν κατά κάποιο τρόπο με τις στοχοθετημένες μεταβλητές του προβλήματος. Για παράδειγμα, αν οι μαθητές θέλουν να συγκρίνουν την περίοδο δυο εκκρεμών και στην εικόνα το ένα σώμα έχει σχεδιαστεί με μεγαλύτερες διαστάσεις λόγω του ότι έχει μεγαλύτερη μάζα, ενδεχομένως κάποιοι από αυτούς να αντιληφθούν γρήγορα τη διαφορά στη μάζα και να χτίσουν ένα επιχείρημα το οποίο να υποστηρίζει την ιδέα ότι η μάζα επηρεάζει την περίοδο του εκκρεμούς (κάτι που είναι επιστημονικά μη ορθό).

### Ευρετική της διαθεσιμότητας

Η έννοια της διαθεσιμότητας, ή διαφορετικά «τι έρχεται πρώτα στο μυαλό», έχει διαδραματίσει σημαντικό ρόλο στην ψυχολογική έρευνα τα τελευταία 50 χρόνια. Σύμφωνα με αυτό το μοντέλο η γνώση που είναι "διαθέσιμη" σε ένα άτομο (π.χ. αποθηκευμένη στη μνήμη) είναι "προσβάσιμη" αν ενεργοποιηθεί με κάποιο τρόπο από το πλαίσιο και, συνεπώς, χρησιμοποιείται για τη λήψη αποφάσεων. Ποια στοιχεία είναι προσβάσιμα μπορούν να έχουν σημαντική επίδραση στις αποφάσεις που λαμβάνει ένα άτομο (Morewedge & Kahneman,2010).

Και οι δύο έννοιες αναφέρονται στην ευκολία ή τη δυσκολία με την οποία οι πληροφορίες έρχονται στο μυαλό. Ωστόσο, ενώ η γνωστική ευκολία σχετίζεται με την ταχύτητα επεξεργασίας (π.χ. το ύψος μπορεί να επεξεργαστεί πιο γρήγορα από την κλίση), η διαθεσιμότητα αναφέρεται στην πιθανότητα μια πληροφορία να έρθει στο μυαλό. Για παράδειγμα, αν ζητηθεί από τους μαθητές να εξηγήσουν για ποιους λόγους 2 εκκρεμή έχουν διαφορετική περίοδο, τότε κάποιοι μαθητές να διατυπώσουν ότι ο ένας λόγος είναι η διαφορετική μάζα των δύο εκκρεμών, κρίση που ενδεχομένως να επηρεάστηκε από το γεγονός ότι η μάζα ήταν μία από τις έννοιες που ήταν διαθέσιμες στο νου των μαθητών.

### Υποκατάσταση ερωτήσεων

Οι άνθρωποι έχουν τη δυνατότητα να παράγουν γρήγορα διαισθητικές απαντήσεις σε δύσκολες ερωτήσεις και πολύπλοκα θέματα. Ένας προτεινόμενος μηχανισμός εξήγησης είναι ο εξής: Αν σε μια δύσκολη ερώτηση δεν βρεθεί γρήγορα μια ικανοποιητική απάντηση το Σύστημα 1 θα ανακαλύψει μια ευκολότερη σχετική ερώτηση και θα απαντήσει σε αυτή. Η λειτουργία απάντησης σε μια ερώτηση στη θέση κάποιας άλλης ονομάζεται υποκατάσταση. Σύμφωνα με τον Kahneman (2011) ερώτηση – στόχος είναι η εκτίμηση που το άτομο σκοπεύει να παράγει, ενώ ευρετική ερώτηση είναι η απλούστερη ερώτηση στην οποία απαντάει αντί της ερώτησης – στόχος. Για παράδειγμα, όταν οι μαθητές ερωτούνται ποια είναι απόσταση μεταξύ του σημείου βολής ενός βλήματος και του σημείου πρόσκρουσης με το έδαφος, κάποιοι μαθητές ενδεχομένως να υποκαταστήσουν την πιο πάνω ερώτηση – στόχο με την εξής ευρετική ερώτηση: Ποιο είναι το βεληνεκές του βλήματος.

### Συνειρμική ενεργοποίηση

Η συνειρμική επεξεργασία χρησιμοποιεί συνδεδεμένες δομές του νου προκειμένου να συμπληρώσει πληροφορίες γρήγορα και αυτόματα σε καταστάσεις που μοιάζουν με προηγούμενες εμπειρίες ή παρατηρήσεις. Γενικά, η κρίση και η λήψη αποφάσεων περιλαμβάνει τη ζύγιση ποικίλων στοιχείων πληροφοριών. Οι γνωστικές μεροληψίες προκύπτουν όταν σε ορισμένες πτυχές των πληροφοριών δίδεται μεγαλύτερο βάρος συστηματικά, ενώ άλλες υποβαθμίζονται ή παραμελούνται, σε σχέση με ένα καθιερωμένο κριτήριο ορθής κρίσης. Η έρευνα υποδεικνύει ότι η έντονα ενεργοποιημένες πληροφορίες είναι πιθανόν να έχουν μεγαλύτερη βαρύτητα από ότι πρέπει, ενώ αντίθετα σχετικές πληροφορίες που είναι ασθενώς ενεργοποιημένες ή δεν έχουν ενεργοποιηθεί καθόλου, τελικά θα παραμεληθούν από τον νου. Σε γενικές γραμμές, η λήψη αποφάσεων βασίζεται σε υπάρχουσες και "επί τόπου" συσχετίσεις μεταξύ των πληροφοριών που έχουν ενεργοποιηθεί. Για παράδειγμα, πολλοί μαθητές συνδέουν τα προβλήματα κρούσης με την εφαρμογή της Αρχής Διατήρησης της Ορμής, ενώ υπάρχουν περιπτώσεις κρούσεων που δεν ισχύει η εν λόγω αρχή (Morewedge & Kahneman,2010).

Οι πιο πάνω ευρετικές διαδικασίες βασίζονται σε συνειρμικές διαδικασίες και πολλές φορές δεν μπορούμε να διακρίνουμε μία τέτοια ευρετική χωρίς να διακρίνουμε και κάποιες από τις υπόλοιπες. Υπάρχουν πολλές περιπτώσεις που οι ευρετικές αυτές εμφανίζονται συνδυαστικά κατασκευάζονται τελικά το πρώτο διαθέσιμο διαισθητικό μοντέλο ερμηνείας μιας δοθείσας κατάστασης. Στη συνέχεια, παρουσιάζονται ευρετικές η οποίες βασίζονται στην επαγωγική παραγωγή κρίσεων.

### Λήψη αποφάσεων βασισμένοι σε μια μόνο αιτιολογία

Σε γενικές γραμμές, οι άνθρωποι απλοποιούν τη συλλογιστική τους χρησιμοποιώντας ένα μοναδικό στοιχείο ή παράγοντα, συχνά το πρώτο που μπορεί να χρησιμοποιηθεί για να δώσει μια εύλογη απάντηση. Αυτή η τάση για απλοποίηση των συλλογισμών επικεντρώνοντας την προσοχή σε μία μεταβλητή και μόνο παραμελώντας άλλες μεταβλητές που επηρεάζουν πιθανώς το πρόβλημα έχει υπογραμμισθεί και από πολλούς ερευνητές στο τομέα της εκπαιδευτικής έρευνας των φυσικών επιστημών. Για παράδειγμα, όταν ζητείται από τους μαθητές να συγκρίνουν τις κινητικές ενέργειες δυο σωμάτων, τείνουν να επικεντρώνουν την προσοχή τους μόνο στην ταχύτητα των σωμάτων και όχι στη μάζα τους (Todd & Gigerenzer,2000).

### Αναγνώριση

Κατά τη λήψη αποφάσεων, οι άνθρωποι συχνά βασίζονται στις πληροφορίες που είναι πιο εύκολο να ανακτηθούν από τη μνήμη, είτε επειδή επεξεργάζονται γρήγορα, είτε επειδή έχουν ενισχυθεί με συχνή έκθεση. Αν ένα από τα πολλά αντικείμενα που έχουμε στη διάθεση μας αναγνωριστεί ενώ τα υπόλοιπα όχι, τότε συμπεραίνουμε ότι το αναγνωρισμένο αντικείμενο έχει την μεγαλύτερο βάρος σε σχέση με τα υπόλοιπα, άρα παίζει πιο σημαντικό ρόλο στη διαμόρφωση της κρίσης μας. Το εν λόγω αντικείμενο παίζει το ρόλο της άγκυρας γύρω από το οποίο θα κινηθούν όλοι οι συλλογισμοί μας. Για παράδειγμα, πολλοί μαθητές όταν θέλουν να συγκρίνουν την ταχύτητα διάδοσης 2 διαφορετικών, ως προς τη συχνότητα, κυμάτων στο ίδιο υλικό, ενδέχεται να δώσουν μεγαλύτερη βαρύτητα στην έννοια της συχνότητας ή του μήκους κύματος μια και εκτίθενται περισσότερο σε αυτές τις έννοιες και όχι στο υλικό διάδοσης που είναι ο μόνος παράγοντας που επηρεάζει την ταχύτητα διάδοσης ενός κύματος(Goldstein & Gigerenzer,2002).

### Επιφανειακή Ομοιότητα

Οι άνθρωποι θεωρούν πως αντικείμενα ή καταστάσεις που μοιάζουν μεταξύ τους με την πρώτη ματιά, ανήκουν στην ίδια κατηγορία και επομένως έχουν παρόμοιες ιδιότητες, συμπεριφορές και εσωτερική δομή. Για παράδειγμα όταν παρουσιάζονται δυο οριζόντιοι ταλαντωτές (σύστημα ελατηρίου – μάζας) που στην μία περίπτωση υπάρχει τριβή και στην άλλη δεν υπάρχει, τότε οι μαθητές τείνουν να θεωρήσουν πως οι ταλαντωτές αυτοί έχουν τις ίδιες ιδιότητες επειδή μοιάζουν με την πρώτη ματιά (Read & Grushka-Cockayne,2011).

### **Επίλογος**

Το έναυσμα για αυτήν την έρευνα ήταν η παρατήρηση ότι οι μαθητές που έχουν τις κατάλληλες γνώσεις και δεξιότητες σε ένα θέμα θετικών επιστημών αδυνατούν, σε πολλές περιπτώσεις, να τις εφαρμόσουν με συνέπεια για να δημιουργήσουν επιστημονικά σωστά συμπεράσματα και λύσεις. Αυτές οι ασυνέπειες είναι επίμονες σε ερωτήσεις που προκαλούν ισχυρές διαισθητικές απαντήσεις. Η παρούσα μελέτη είχε ως στόχο να εντοπίσει τους γνωστικούς μηχανισμούς που συμβάλλουν σε αυτές τις ασυνέπειες. Σύμφωνα με τις θεωρίες περί των δύο συστημάτων του νου, οι μαθητές βασίζονται σε πολλές ευρετικές για να λάβουν αποφάσεις και να απαντήσουν σε μια δεδομένη ερώτηση. Πολλές από τις ευρετικές που περιγράφονται σε αυτήν την έρευνα ενεργούν σε συνδυασμό όταν οι μαθητές εργάζονται πάνω στη λύση ενός προβλήματος που τους δίνεται. Αυτή η μεταξύ τους διασύνδεση καθιστά δύσκολη την πρόταση στοχευμένων στρατηγικών που μπορούν να βοηθήσουν τους μαθητές να αποφύγουν αυτές τις γνωστικές προκαταλήψεις. Τέλος, χρειάζεται πρόσθετη έρευνα από την επιστημονική κοινότητα προκειμένου να διερευνηθούν περεταίρω οι ερευτικές διαδικασίες και πως αυτές χρησιμοποιούνται κατά τη διάρκεια της επίλυσης των προβλημάτων στις θετικές επιστήμες προκειμένου να σχεδιαστούν προγράμματα σπουδών που να βοηθούν τους μαθητές να αναγνωρίζουν τις ευρετικές αυτές και τελικά να τις αποφεύγουν. Με αυτόν τον τρόπο οι μαθητές μπορούν να βελτιώσουν τόσο την επίδοσή τους στις φυσικές επιστήμες, όσο τις αποφάσεις που λαμβάνουν στη ζωή τους.

**Βιβλιογραφική Αναφορά**

Evans, J. S. B. (2006). The heuristic-analytic theory of reasoning: Extension and evaluation. *Psychonomic Bulletin & Review*, *13*(3), 378-395.

Frankish, K. (2010). Dual-process and dual-system theories of reasoning. *Philosophy Compass*, *5*(10), 914-926.

Gette, C. R., Kryjevskaia, M., Stetzer, M. R., & Heron, P. R. (2018). Probing student reasoning approaches through the lens of dual-process theories: A case study in buoyancy. Physical Review Physics Education Research, 14(1), 010113.

Gilovich, T., Griffin, D., & Kahneman, D. (Eds.). (2002). *Heuristics and biases: The psychology of intuitive judgment*. Cambridge university press.

Goldstein, D. G., & Gigerenzer, G. (2002). Models of ecological rationality: the recognition heuristic. *Psychological review*, *109*(1), 75.

Heckler, A. F. (2011). 8 The Ubiquitous Patterns of Incorrect Answers to Science Questions: The Role of Automatic, Bottom-up Processes. *Psychology of Learning and Motivation-Advances in Research and Theory*, *55*, 227.

Johnson-Laird, P. N. (2006). *How we reason*. Oxford University Press, USA.

Kahneman, D. (2011). *Thinking, fast and slow* (Vol. 1). New York: Farrar, Straus and Giroux.

Kryjevskaia, M., Stetzer, M. R., & Heron, P. R. (2012). Student understanding of wave behavior at a boundary: The relationships among wavelength, propagation speed, and frequency. *American Journal of Physics*, *80*(4), 339-347.

Kryjevskaia, M., Stetzer, M. R., & Grosz, N. (2014). Answer first: Applying the heuristic-analytic theory of reasoning to examine student intuitive thinking in the context of physics. *Physical Review Special Topics-Physics Education Research*, *10*(2), 020109.

McClary, Lakeisha, and Vicente Talanquer. "Heuristic reasoning in chemistry: Making decisions about acid strength." International Journal of Science Education 33.10 (2011): 1433-1454.

Meltzer, D. E., & Thornton, R. K. (2012). Resource letter ALIP–1: active-learning instruction in physics. *American journal of physics*, *80*(6), 478-496.

Morewedge, C. K., & Kahneman, D. (2010). Associative processes in intuitive judgment. *Trends in cognitive sciences*, *14*(10), 435-440.

Oppenheimer, D. M. (2008). The secret life of fluency. *Trends in cognitive sciences*, *12*(6), 237-241.

Read, D., & Grushka-Cockayne, Y. (2011). The similarity heuristic. *Journal of Behavioral Decision Making*, *24*(1), 23-46.

Shah, A. K., & Oppenheimer, D. M. (2008). Heuristics made easy: An effort-reduction framework. *Psychological bulletin*, *134*(2), 207.

Talanquer, V. (2014). Chemistry education: Ten heuristics to tame. *Journal of Chemical Education*, *91*(8), 1091-1097.

Thompson, V. A., Turner, J. A. P., & Pennycook, G. (2011). Intuition, reason, and metacognition. *Cognitive psychology*, *63*(3), 107-140.

Thompson, V. A., Evans, J. S. B., & Campbell, J. I. (2018). Matching bias on the selection task: It's fast and feels good. In *New Paradigm Psychology of Reasoning* (pp. 194-215). Routledge.

Todd, P. M., & Gigerenzer, G. (2000). Précis of simple heuristics that make us smart. *Behavioral and brain sciences*, *23*(5), 727-741.

Tversky, A., & Kahneman, D. (1973). Availability: A heuristic for judging frequency and probability. *Cognitive psychology*, *5*(2), 207-232.

Wood, A. K., Galloway, R. K., & Hardy, J. (2016). Can dual processing theory explain physics students' performance on the Force Concept Inventory?. Physical Review Physics Education Research, 12(2), 023101.